\documentclass[aps,pra,twocolumn,groupedaddress]{revtex4-1}
\usepackage{bm}
\usepackage[dvipdfmx]{graphicx}
\usepackage{booktabs}
\usepackage{amsmath}

\begin{document}

%Title of paper
\title{A Three-Dimensional Optical Lattice of Ytterbium and Lithium Atomic Mixture}

\author{Hideaki Hara,$^1$ Hideki Konishi,$^1$ Shuta Nakajima,$^1$ Yosuke Takasu,$^1$ Yoshiro Takahashi$^{1,2}$}
\affiliation{$^1$Department of Physics, Graduate School of Science, Kyoto University, Kyoto 606-8502, Japan\\
$^2$CREST, Japan Science and Technology Agency, Chiyoda-ku, Tokyo 102-0075, Japan
}

\date{\today}

\begin{abstract}
We develop an optical lattice system for an ultracold atomic gas mixture of ytterbium ($^{174}$Yb) and lithium ($^6$Li), 
which is an ideal system to study disorder and impurity problems.
We load a Bose-Einstein condensate of $^{174}$Yb into a three-dimensional optical lattice 
and observe the interference patterns in time-of-flight (TOF) images.
Furthermore, we perform a laser spectroscopy of $^{174}$Yb in an optical lattice using the ultra-narrow optical transition $^1$S$_0$-$^3$P$_2$ 
in both cases with and without $^6$Li.
Due to the weak interspecies interaction, 
we do not observe the clear influences of $^6$Li in the obtained interference patterns and the excitation spectra.
However, this is an important first step of optical control of atomic impurity in ultracold fermions.
We also measure the polarizabilities of the $^3$P$_2$ state of  $^{174}$Yb atoms in an optical trap with a wavelength of 1070 nm.
We reveal that 
the polarizability can be tuned to positive, zero, or the same as the ground state, which are useful for certain applications.
\end{abstract}

\maketitle

\section{Introduction}

\label{intro}
Over the past decade, a considerable number of theoretical and experimental studies 
on ultracold atomic gases in an optical lattice have been conducted\cite{BlochRMP}.
The pioneering experimental research is the observation of the superfluid to Mott-insulator transition 
by loading a Bose-Einstein condensate (BEC) into an optical lattice \cite{Mott}.
Femiomic atoms in an optical lattice are also important in condensed matter physics 
since they directly correspond to interacting electrons in solids.
Fermions in an optical lattice could be employed for studies of high-$T_{c}$ superconductivity, quantum magnetism, and so on.
As groundbreaking experiments, fermionic Mott insulators are also realized \cite{FermiMottEsslinger,FermiMottBloch}.
One of the advantages of the ultracold atoms in an optical lattice over solid state systems is high controllability of Hubbard parameters 
such as the on-site interaction and the tunneling rate between neighboring sites.
Another advantage is the fact that defects of optical lattices can be controlled, 
i.e., even disorder and impurity are treated as controllable parameters.
They are thought to play a key role in high-$T_c$ superconductivity, Anderson localization, 
and some novel quantum phases such as Bose glass etc.
Controlled disorder systems in ultracold gases are so far provided 
by an optical speckle pattern \cite{Anderson1,Anderson2,AndersonFermion}, 
an incommensurate optical lattice potential superimposed on the main lattice \cite{BoseGlass}, 
or by loading two atomic species into an optical lattice \cite{BFLattice1,BFLattice2}.
In particular, ultracold gas mixtures in optical lattices are well described by the binary-alloy Anderson-Hubbard model \cite{binary-alloy}, 
where local disorders are given by random on-site energies originating from interspecies interactions between localized impurities and mobile atomic species.

\begin{figure}
\centering
\includegraphics[width=6cm,keepaspectratio]{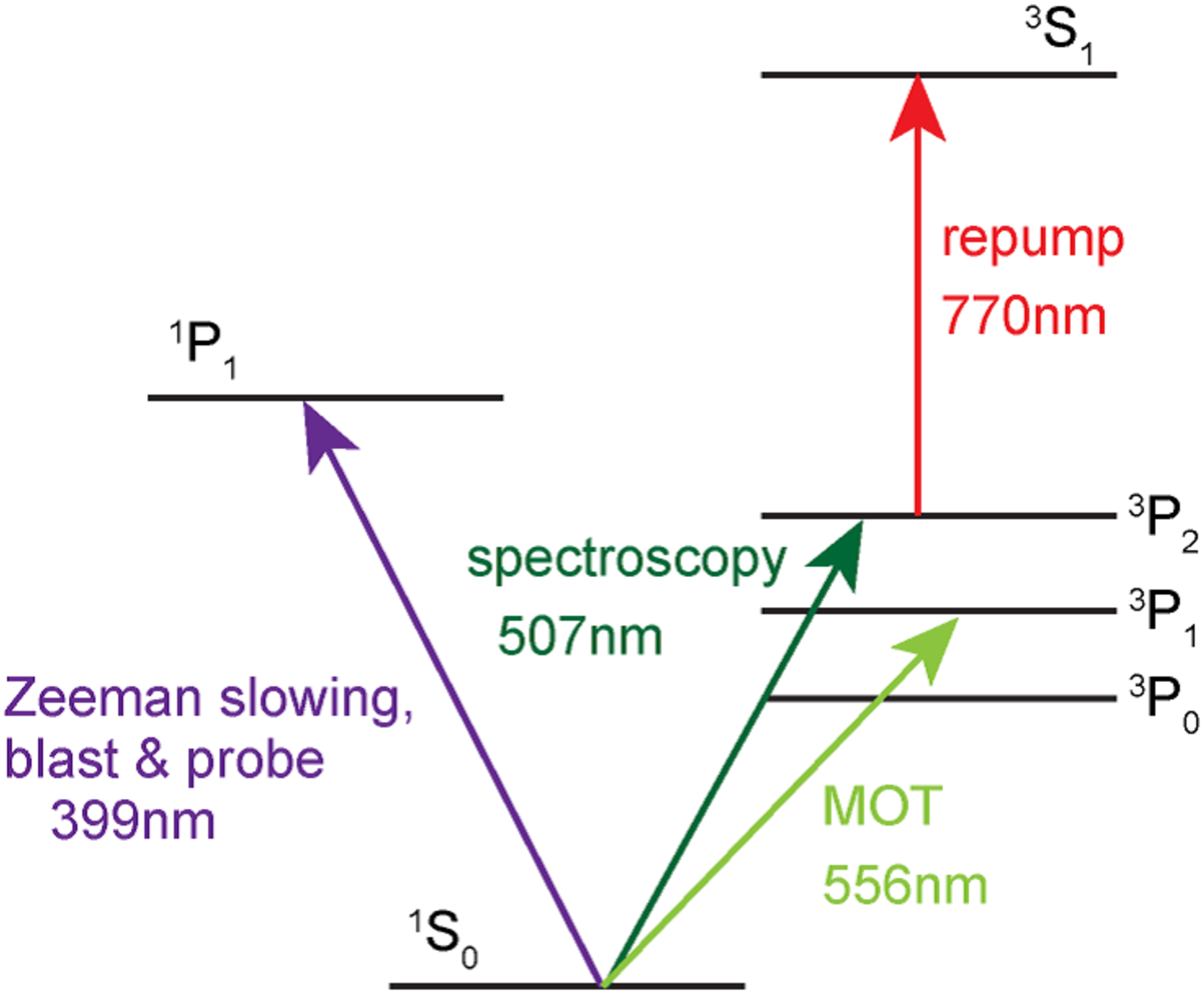}
\caption{(Color online)
Energy diagram of $^{174}$Yb.
The $^1$S$_0$-$^1$P$_1$ transition ($\lambda$ = 399 nm) is used 
for Zeeman slowing, for absorption imaging, and for removing the $^1$S$_0$ state atoms remaining in a trap (for blast).
The $^1$S$_0$-$^3$P$_1$ transition ($\lambda$ = 556 nm) is used for MOT.
The ultra-narrow $^{1}$S$_0$-$^3$P$_2$ transition ($\lambda$ = 507 nm) is used for the spectroscopy.
The $^3$P$_2$-$^3$S$_1$ transition ($\lambda$ = 770 nm) is used for repumping the $^3$P$_2$ atoms back to the $^1$S$_0$ state via the $^3$P$_1$ state.}
\label{fig:energydiagram}
\end{figure}

We focus on a mixture of $^{174}$Yb and $^6$Li atoms in an optical lattice system with an interest in disorder and impurity problems.
The most distinctive feature of this mixture is the very large mass ratio of 29.
Because of this large mass ratio, 
the tunneling rate of $^{174}$Yb atoms can be more than three orders of magnitude smaller than that of $^{6}$Li atoms 
in a deep optical lattice with a wavelength of 1064 nm, for example.
Therefore, heavy atoms of $^{174}$Yb play a role of localized impurities and light atoms of $^6$Li behave as itinerant electrons.
With regard to the interspecies interaction, 
the absolute values of $s$-wave scattering length have already been measured \cite{QDYbLi,Gupta}.
It is desirable to resonantly control interspecies interactions, but it is pointed out by a theoretical calculation 
that precise control of  the $s$-wave scattering length between $^{174}$Yb and $^6$Li is experimentally challenging 
due to the extremely narrow width of Feshbach resonances \cite{YbLiFeshbach}.
On the other hand, if the metastable $^3$P$_2$ state of $^{174}$Yb atoms, 
whose energy diagram is shown in Fig. \ref{fig:energydiagram}, 
are introduced as localized impurities instead of the ground $^1$S$_0$ state atoms, 
there is a possibility of controlling the interaction between the impurity and the itinerant $^6$Li atoms in the ground $^2$S$_{1/2}$ state 
via anisotropy-induced Feshbach resonances \cite{AnisotropyInduced,3P2Feshbach,3P2-Li-Feshbach}, 
which allows us to suddenly switch from a weak to a strong interaction.
In particular, the ultra-narrow optical transition $^1$S$_0$-$^3$P$_2$ in the presence of the Feshbach resonances 
enables the exploration of impurity problems such as Anderson's Orthogonality Catastrophe, 
in which impurities are localized by an optical lattice potential immersed in a sea of itinerant fermions \cite{Anderson'sOC}.
Note that the study of the dynamics is possible in this system, differently from a solid-state system with the fast time scale, 
and our system could be a candidate to study the Anderson's Orthogonality Catastrophe 
not only in frequency domain but also in time domain.
In addition, a molecule produced from this mixture has not only an electric dipole moment but also electronic spin degrees of freedom in the ground state.
Spin-doublet molecules in an optical lattice enables us to implement quantum simulation of lattice-spin models \cite{latticespinmodel}.%\cite{latticespinmodelrealizing}.

In this paper, we report the first realization of an optical lattice for the mixture of $^{174}$Yb and $^6$Li.
We successfully load $^{174}$Yb BEC with Fermi degenerate $^6$Li into a three-dimensional (3D) optical lattice with the wavelength of 1064 nm.
We measure the influence of $^6$Li atoms on the coherence property of $^{174}$Yb and on the excitation spectra of $^{174}$Yb 
using the ultra-narrow optical transition $^1$S$_0$-$^3$P$_2$.
In addition, we measure polarizabilities of the $^3$P$_2$ state $^{174}$Yb in a laser field with a wavelength of 1070 nm 
in order to reveal whether they can be trapped in a 1 $\mu$m optical trap.

\section{Experimental Setup}
\label{sec:Experimental setup}

\subsection{Preparation of a quantum degenerate mixture of $^{174}$Yb and $^{6}$Li}
\label{subsec:preparation}
We first prepare a quantum degenerate mixture of bosonic $^{174}$Yb and fermionic $^{6}$Li in an optical trap.
The basic scheme is the same as that described previously \cite{QDYbLi,Okano}.
The experiment begins with a simultaneous magneto-optical trap (MOT).
For $^{174}$Yb, the $^1$S$_0$-$^1$P$_1$ transition ($\lambda$ = 399 nm) is used for Zeeman slowing and probing.
The $^1$S$_0$-$^3$P$_1$ intercombination transition ($\lambda$ = 556 nm) is used 
for magneto-optical trapping.
For $^6$Li, the $^2$S$_{1/2}$-$^2$P$_{3/2}$ (D$_2$) transition ($\lambda$ = 671 nm) is used for Zeeman slowing, 
magneto-optical trapping, and probing.
In order to realize a mixture of $^{174}$Yb BEC and degenerate $^6$Li, 80-120 s is needed for loading of $^{174}$Yb atoms in a MOT, 
followed by 5-10 s for loading of $^{6}$Li.
We transfer both atoms from the MOT into a crossed optical far-off-resonance trap (FORT) and 
perform sympathetic cooling of $^6$Li with evaporatively cooled $^{174}$Yb.
The horizontal FORT beam ($\lambda$ = 1070 nm) is in an ellipsoidal shape, 
with the beam waists of 106 $\mu$m along the horizontal direction and 22 $\mu$m along the vertical direction, 
in order to obtain a large trap volume and suppress the gravitational sag at the final stage of evaporative cooling.
Another FORT beam ($\lambda$ =1083 nm), propagating along the almost vertical direction, 
has a round shape with a beam waist of 104 $\mu$m.
After evaporative cooling for 7 s, we obtain a quantum degenerate mixture composed of an almost pure $^{174}$Yb BEC and a degenerate Fermi gas of $^{6}$Li.
The typical atom number of $^{174}$Yb BEC is $N_{\mathrm{BEC}}=(7.5\pm2.5)\times10^{4}$.
For $^6$Li, the number is $N_{\mathrm{Li}}=(1.5\pm0.5)\times10^{4}$, 
and the temperature is $T_{\mathrm{Li}}=290$ nK and $T/T_{F}\simeq0.2$.
The trap frequencies of $^{174}$Yb at the end of evaporation are $2\pi\times(47,64,130)$ Hz 
and those of $^6$Li are $2\pi\times(354,479,1522)$ Hz.
In this situation, the $^{174}$Yb BEC cloud is located lower than that of Fermi degenerate $^6$Li by 7.8 $\mu$m.
The radii of the atom cloud along the vertical direction are 2.9 $\mu$m for $^{174}$Yb and 5.7 $\mu$m for $^6$Li, which results in 
about 15 \% of $^{174}$Yb BEC cloud length spatially overlapped with the $^6$Li cloud.

\subsection{An optical lattice}
\begin{figure}
\centering
\includegraphics[width=7cm,keepaspectratio]{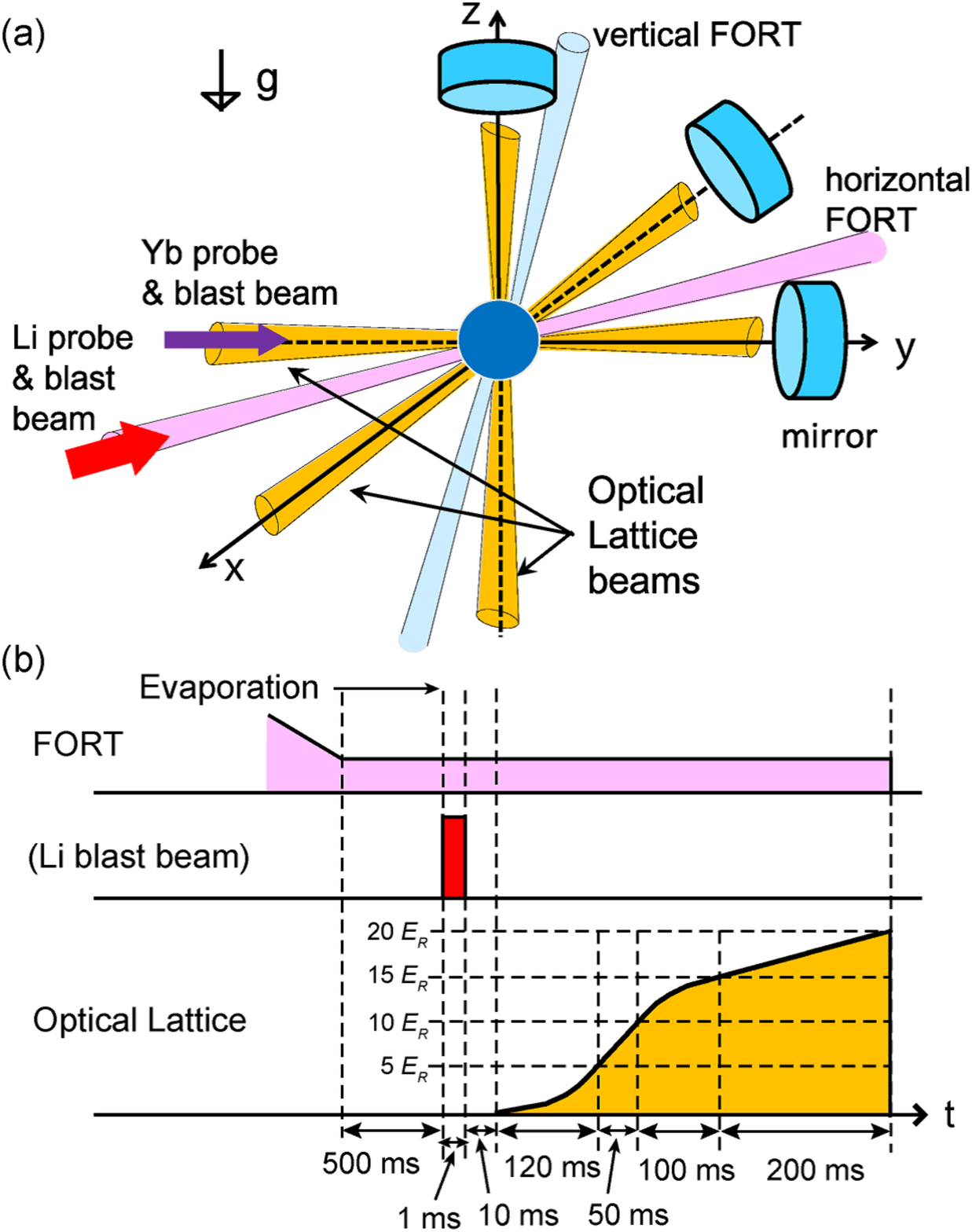}
\caption{(Color online)
Experimental setup (a) and procedure (b) to load atoms into an optical lattice.
For the data of pure $^{174}$Yb, we remove $^6$Li atoms by a Li blast laser.
For the mixture, we wait for the same duration without irradiating the Li blast laser.
$E_{R}$ in the timing chart denotes the recoil energy of $^{174}$Yb.}
\label{fig:setuptimechart}
\end{figure}
At the wavelength $\lambda_{L}$ = 1064 nm of our optical lattice beams, 
the ratio of the trap depths for $^{174}$Yb ($V_{\mathrm{Yb}}$) and $^6$Li ($V_{\mathrm{Li}}$) is 0.50.
Due to the large mass ratio of 29, however, we have $s_{\mathrm{Yb}}/s_{\mathrm{Li}} = 14.5$, 
where $s_{\mathrm{Yb}}=V_{\mathrm{Yb}}/E_{R}^{\mathrm{Yb}}$ and $s_{\mathrm{Li}}=V_{\mathrm{Li}}/E_{R}^{\mathrm{Li}}$ and $E_{R} = \hbar^{2}k_{L}^{2}/2m$ is the recoil energy 
with $k_{L} = 2\pi/\lambda_{L}$ and the atomic mass $m$.
Optical lattice potentials for the same $s$ have the same band structures and the same Wannier functions 
where energy scales are different due to $E_{R}$.
The recoil energy of $^{174}$Yb ($E_{R}^{\mathrm{Yb}}$) is 49 nK 
and that of $^{6}$Li ($E_{R}^{\mathrm{Li}}$) is 1.4 $\mu$K.
As a result, the tunneling rate of $^6$Li is 
more than three orders of magnitude larger than that of $^{174}$Yb at $s_{\mathrm{Yb}}=20$, for example.

The experimental setup is shown in Fig. \ref{fig:setuptimechart} (a).
The lattice potential is formed by three orthogonal retroreflected laser beams, 
which consist of two horizontally propagating beams (x, y-lattice) and a vertically propagating beam (z-lattice).
The angle between the horizontal FORT beam and x,y-lattice beams is $45^{\circ}$.
The lattice beams are created by a fiber laser amplifier which is seeded by a Nd: YAG laser 
with a  linewidth of less than 10 kHz.
In order to eliminate undesirable interferences between each beam, 
the frequencies of the three lattice beams are shifted by 5 MHz relative to each other 
and the polarization is orthogonal to each other.
Acousto-optic modulators (AOMs) are used for frequency shifts and for power control.
The transverse modes of the three lattice beams are cleaned by single-mode optical fibers.
The waist sizes of the lattice beams are measured with a beam profiler 
to be 81, 88, and 98 $\mu$m for x, y, and z-lattice, respectively.
The power is monitored by the weakly transmitted beams through the retrorefrecting mirrors and 
the intensity is stabilized by using the AOMs.
The lattice potential depth is calibrated by observing the oscillating period of the interference by a pulsed optical lattice for $^{174}$Yb BEC \cite{pulsedlattice}.

\section{Loading into a 3D Optical Lattice}
\begin{figure}%[h!]
\centering
\includegraphics[width=7cm,keepaspectratio]{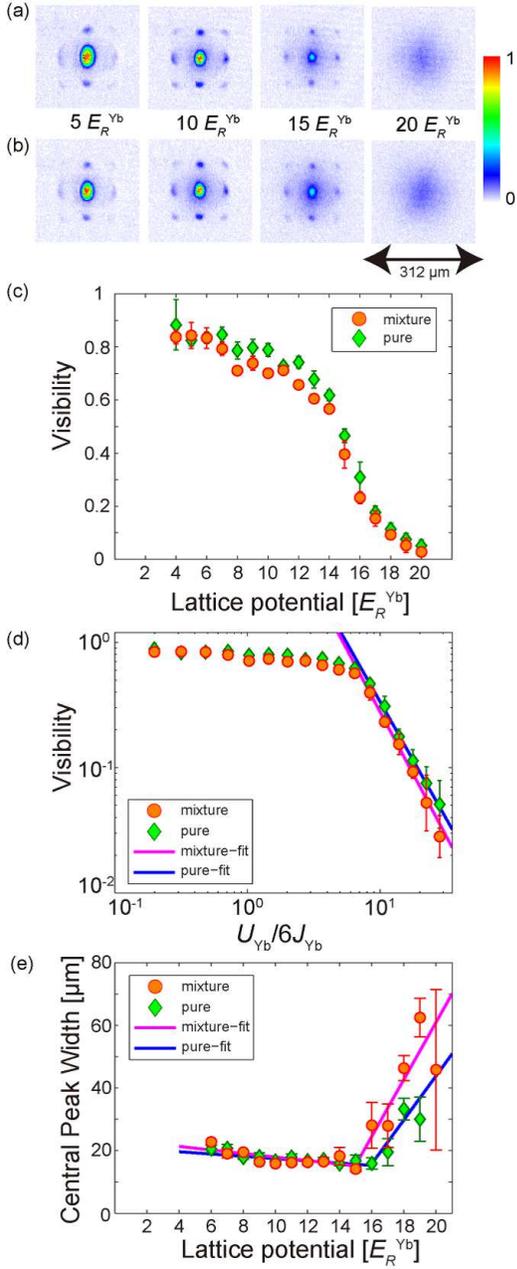} %2column
\caption{(Color online)
TOF absorption images of multiple matter-wave interference patterns of $^{174}$Yb (a) without $^6$Li and (b) with $^6$Li.
The lattice potential depths are 5 $E_{R}^{\mathrm{Yb}}$, 10 $E_{R}^{\mathrm{Yb}}$, 15 $E_{R}^{\mathrm{Yb}}$, and 20 $E_{R}^{\mathrm{Yb}}$ from left to right.
The expansion time is 17 ms.
These images are average over 4 $\sim$ 6 measurements.
(c) Visibility of $^{174}$Yb for the mixture (red circle) and the pure $^{174}$Yb (green diamond) as a function of the lattice potential depth.
Each data point is average over 4 $\sim$ 6 measurements and the error bars indicate standard deviations.
(d) Visibility versus $U/zJ$ in a log-log plot.
The lines are linear fits to the data in the range from 15 to 20 $E_{R}^{\mathrm{Yb}}$.
The slope of the linear fit $\beta$ for the mixture is $-1.99\pm0.09$ and slightly lower than that for the pure $^{174}$Yb $-1.87\pm0.04$.
(e) Central interference peak width of $^{174}$Yb for the mixture (red circle) and the pure $^{174}$Yb (green diamond) 
as a function of the lattice depth.
The lines are the fits to two linear functions to determine the critical depths.
The intersections are $(15.0\pm5.0)$ $E_{R}^{\mathrm{Yb}}$ for the mixture and $(15.9\pm7.3)$ $E_{R}^{\mathrm{Yb}}$ for the pure $^{174}$Yb.
They are almost identical within the error bars.}
\label{fig:visibility}
\end{figure}

We load the atomic mixture of $^{174}$Yb BEC and a degenerate Fermi gas of $^6$Li into the 3D optical lattice.
While in our target experiments of impurity problems $^{174}$Yb atoms should play a role of localized impurities, 
its first step of the successful formation of the optical lattice is easily confirmed 
by measuring the matter-wave interference patterns of $^{174}$Yb BEC at a relatively shallow optical lattice 
where the phase coherence exists at the superfluid state of $^{174}$Yb atoms.
The experimental sequence is shown in Fig. \ref{fig:setuptimechart} (b).
After the preparation of a quantum degenerate mixture the lattice potential depth is gradually increased with keeping the FORT potential constant.
We take TOF images of $^{174}$Yb atoms with various lattice depths, as shown in  Fig. \ref{fig:visibility} (a).
When the lattice depth is shallow, 
multiple matter-wave interference patterns appear in the TOF images, 
which signals that $^{174}$Yb atoms delocalize over the lattice sites and there exists phase coherence.
As increasing the lattice depth, such an interference pattern disappears gradually, 
which indicates that atoms localize and there is no longer phase coherence.
In this measurement, the lattice depth for $^{174}$Yb changes up to 20 $E_{R}^{\mathrm{Yb}}$.
In the presence of a weak harmonic confining potential, a Mott insulator in an optical lattice has a shell structure.
The estimation in the atomic limit $U/zJ\to\infty$ \cite{atomiclimit} 
gives the occupation number at the center of the trap of 11 for 20 $E_{R}^{\mathrm{Yb}}$.
Here $U$ is the on-site interaction energy between bosonic atoms, $J$ is the tunneling matrix element, 
and $z=6$ is the number of nearest neighboring sites in a 3D cubic lattice.
The typical number of $^{174}$Yb atoms is $7.5\times10^4$ 
and the trap frequencies of the confinement potential for $^{174}$Yb are $2\pi\times(62,75,136)$ Hz at that lattice depth.

To obtain quantitative information from the interference patterns of TOF images, we evaluate the visibility and the width of the central interference peak.
The visibility is defined as 
$\mathcal{V}=(N_{\mathrm{max}}-N_{\mathrm{min}})/(N_{\mathrm{max}}+N_{\mathrm{min}})$  \cite{visibility}.
Here $N_{\mathrm{max}}$ is the sum of the number of atoms in the first order interference peaks 
and $N_{\mathrm{min}}$ is the sum of the number of atoms in the diagonal regions with the same distance from the center peak.
Figure \ref{fig:visibility} (c) shows thus evaluated visibilities for $^{174}$Yb.
According to the mean-field theory for the homogeneous system, 
the critical value of $U/zJ$ for the superfluid to Mott insulator transition with $n_c$ atoms per site is
$(U/zJ)_c=(\sqrt{n_c}+\sqrt{n_{c}+1})^2$ \cite{MottTransition}.
From this formula, the critical values of the lattice depth are 
13.6, 15.6, 17.0, 18.1, 18.9, 19.6, and 20.3 $E_{R}^{\mathrm{Yb}}$ for $n_c$=1, 2, 3, 4, 5, 6, and 7. 
In this estimation we use the approximate expressions $U/E_{R}=5.97(a/\lambda_{L})s^{0.88}$ 
and $J/E_{R}=1.43s^{0.98}\exp(-2.07\sqrt{s})$ \cite{approximateUJ} 
with a scattering length between $^{174}$Yb $a=5.55$ nm \cite{kitagawa}.
For larger values of $U/zJ$, the visibility is $\mathcal{V}\sim\frac{4}{3}(n_{c}+1)(U/zJ)^{-1}$ \cite{visibility}, 
where $n_{c}$ denotes a filling factor in a homogeneous system.
To compare our result with this estimation, we show the visibility as a function of $U/zJ$ in a log-log plot in Fig. \ref{fig:visibility} (d).
The visibility $\mathcal{V}$ starts to decrease at around 15 $E_{R}^{\mathrm{Yb}}$.
This is consistent with the critical values for the transition given above.
The results of the fit to the linear function indicate that $\mathcal{V}\propto(U/zJ)^{\beta}$ 
with $\beta = -1.87 \pm 0.04$.
This value is almost twice as large as the theoretical value of $-1$.
This implicates that there is some mechanism for decrease of the coherence.

The phase coherence of the bosons in an optical lattice is also characterized by the width of the central interference peak.
Figure \ref{fig:visibility} (e) shows the full width at half maximum of the central peak as a function of the lattice depth.
The widths start to increase at around 16 $E_{R}^{\mathrm{Yb}}$.
If the critical value of the lattice depth for the transition is defined as the intersection of the two linear functions, 
it is $(15.9\pm7.3)$ $E_{R}^{\mathrm{Yb}}$ for the $^{174}$Yb, 
which almost agrees with the values estimated from the mean-field approximation.

The above measurements on the $^{174}$Yb matter-wave interference show that 
we could successfully load the $^{174}$Yb atoms into the optical lattice.
It should be also true for the $^6$Li atoms, 
since the 1064 nm lattice laser should produce 
a lattice potential for $^6$Li atoms twice as deep as for $^{174}$Yb atoms 
corresponding to 1.4 $E_{R}^{\mathrm{Li}}$ at the maximal potential depth in this measurement.
Furthermore, we also try to observe the influence of the $^6$Li for the $^{174}$Yb atom signals.
In order to clearly discuss the influence of $^6$Li atoms, 
for the data without $^6$Li atoms we remove $^6$Li atoms in the FORT 
by irradiating a laser pulse resonant to the $^2$S$_{1/2}$-$^2$P$_{3/2}$ (D$_2$) transition (Li blast laser, $\lambda$ = 671 nm) for 1 ms, followed by 10 ms holding time, 
before ramping up the lattice potential as shown in Fig. \ref{fig:setuptimechart} (b).
For the data with $^6$Li atoms, shown in Fig. \ref{fig:visibility} (b), we wait for the same duration without irradiating the blast laser.
In this experiment, the influence of $^6$Li on the phase coherence of $^{174}$Yb can be recognized as shown in Fig. \ref{fig:visibility} (c), (d), and (e), but it is very small.
This is in good contrast with the case of $^{87}$Rb-$^{40}$K mixture \cite{BFLattice1,BFLattice2} 
where even very small fraction of fermionic atoms dramatically changes the coherence property of the bosonic cloud.
We think several reasons for small influence.
The first reason is the imperfect spatial overlap of $^{174}$Yb and $^6$Li.
As mentioned in Sect. \ref{subsec:preparation}, the center of $^{174}$Yb BEC is 7.8 $\mu$m lower than that of $^6$Li at the final phase of evaporation due to the gravitational sag.
After loading of the atoms into the optical lattice, the $^{174}$Yb cloud is broadened to 5.3 $\mu$m radius along the z-direction 
for the case of the adiabatic loading and the atomic limit $U/zJ\to\infty$.
Even in this situation the overlap is up to 30 \%.
The second is that the interspecies interaction is too weak to significantly change the coherence property of $^{174}$Yb.
The ratio of the absolute value of the on-site $^{174}$Yb-$^6$Li interaction $U_{\mathrm{Yb-Li}}$ to that between $^{174}$Yb $U_{\mathrm{Yb}}$ is $|U_{\mathrm{Yb-Li}}|/U_{\mathrm{Yb}}=0.73<1$, 
which supports the small influence.
Here we use 
$U_{BF}=\sqrt{\frac{16}{\pi}}\Big(1+\frac{m_{B}}{m_{F}}\Big)\Big(\frac{1}{\sqrt{s_{B}}}+\frac{1}{\sqrt{s_{F}}}\Big)^{-3/2}k_{L}a_{BF}E_{R}^{B}$ \cite{UBF}.
$E_{R}^{B}$ is the recoil energy of the bosonic atom.
Note that the possibility of phase separation is excluded in this experiment.
A stability condition for a mixture is \cite{BF-phase-separation}
\begin{equation}
\label{eq:stability}
n_{\mathrm{Li}}^{-1/3} \ge 3 \pi \bigg( \frac{g_{s}}{6 \pi^{2}} \bigg)^{2/3} \frac{m_{\mathrm{Yb}} m_{\mathrm{Li}} a_{\mathrm{Yb-Li}}^{2}}{\mu_{\mathrm{Yb-Li}}^{2} a_{\mathrm{Yb}}} ,
\end{equation}
where $n_{\mathrm{Li}}$ is the density of $^6$Li, $g_{s}=2$ is the number of spin components of $^6$Li, 
$m_{\mathrm{Yb}}$ and $m_{\mathrm{Li}}$ are the masses of $^{174}$Yb and $^6$Li, 
$\mu_{\mathrm{Yb-Li}}$ is the reduced mass of $^{174}$Yb and $^6$Li, 
$a_{\mathrm{Yb-Li}}$ is the $s$-wave scattering length between $^{174}$Yb and $^6$Li, 
and $a_{\mathrm{Yb}}$ is that between $^{174}$Yb atoms.
Here we consider a homogeneous system for simplicity.
In our experiment, 
$n_{\mathrm{Li}}^{-1/3} = 0.60$ $\mu$m and 
$3 \pi \Big( \frac{g_{s}}{6 \pi^{2}} \Big)^{2/3} \frac{m_{\mathrm{Yb}} m_{\mathrm{Li}} a_{\mathrm{Yb-Li}}^{2}}{\mu_{\mathrm{Yb-Li}}^{2} a_{\mathrm{Yb}}} = 5.5$ nm, 
which means that the stability condition is satisfied.
Therefore the possibility of phase separation is eliminated.

We also mention the adiabaticity of the atom loading into the optical lattice.
Unfortunately, differently from the lattice experiments for $^{174}$Yb so far realized using a 532 nm  laser \cite{YbLattice}, 
it is rather difficult to adiabatically load the atoms into the optical lattice with 1064 nm.
In fact, we ramp down the potential to 10 $E_{R}^{\mathrm{Yb}}$ 
after ramping up the lattice depth to 20 $E_{R}^{\mathrm{Yb}}$, 
interference patterns are not restored.
It indicates the lack of adiabaticity during ramping up or down the lattice depth.
In addition to the fact that the on-site interaction energy and the tunneling rate for a $\lambda_{L}$ = 1064 nm lattice 
are 1/8 and 1/4 of those for a $\lambda_{L}$ = 532 nm lattice, respectively, 
the difference of the energy offset between the neighboring sites due to an external harmonic confinement for a $\lambda_{L}$ = 1064 nm lattice is 
larger than that for a $\lambda_{L}$ = 532 nm lattice.
Therefore in a $\lambda_{L}$ = 1064 nm lattice atoms are easily localized.
This non-adiabaticity might explain the larger values of $\beta$ in the visibility measurements.

\section{Spectroscopy of $^{174}$Yb-$^6$Li Mixture in an Optical Lattice Using the Ultra-Narrow Optical Transition of $^{174}$Yb Atoms}

\subsection{Polarizability of the $^{3}$P$_{2}$ state of $^{174}$Yb atom in an optical trap with a wavelength of 1070 nm}

We measure the polarizabilities, or AC Stark shifts of the $^3$P$_2$ state of $^{174}$Yb atoms 
by a laser field with a wavelength of 1 $\mu$m.
This information is important because the difference between the polarizabilities of the ground and the $^3$P$_2$ states 
is directly related to spectrum inhomogeneous broadening.
If the polarizability can be set to the same as that of the ground state, like a magic wavelength condition, 
then we can expect a narrow spectrum which is only limited by an atomic interaction.
In the case of the negative polarizability, the excited atoms will escape from the trap or lattice, 
which limits the resolution of the spectroscopy.
If the polarizability can be set to zero, then this situation is ideal for the photo-emission spectroscopy \cite{photoemission}.

The AC Stark shift is described as $U_{\mathrm{dip}}=-\frac{1}{2\epsilon_{0}c} \alpha I$ \cite{GrimmODT}.
Here $\alpha$ is the polarizability and $I$ is the intensity of the laser field.
We determine the poralizability of the $^3$P$_2$ state 
by directly measuring the transition frequency $^1$S$_0$-$^3$P$_2$ ($\lambda$ = 507 nm).
The transition frequency $\nu$ of atoms trapped in an optical trap is 
given by using the bare resonant frequency $\nu_0$ 
and the polarizabilities of the $^1$S$_0$ state ($\alpha_g$) and the $^3$P$_2$ state ($\alpha_e$) as
\begin{equation}
\nu = \nu_{0}-\frac{1}{2\epsilon_{0}ch}(\alpha_{e}-\alpha_{g}) I .
\end{equation}
It is noted that $\alpha_{e}$ depends not only on the laser frequency $\omega$ 
but also on the polarization of the laser field and the magnetic field strength and direction.
In addition, different magnetic sublevels of the $^3$P$_2$ state have different polarizabilities.

\begin{figure}
\centering
\includegraphics[width=8cm,keepaspectratio]{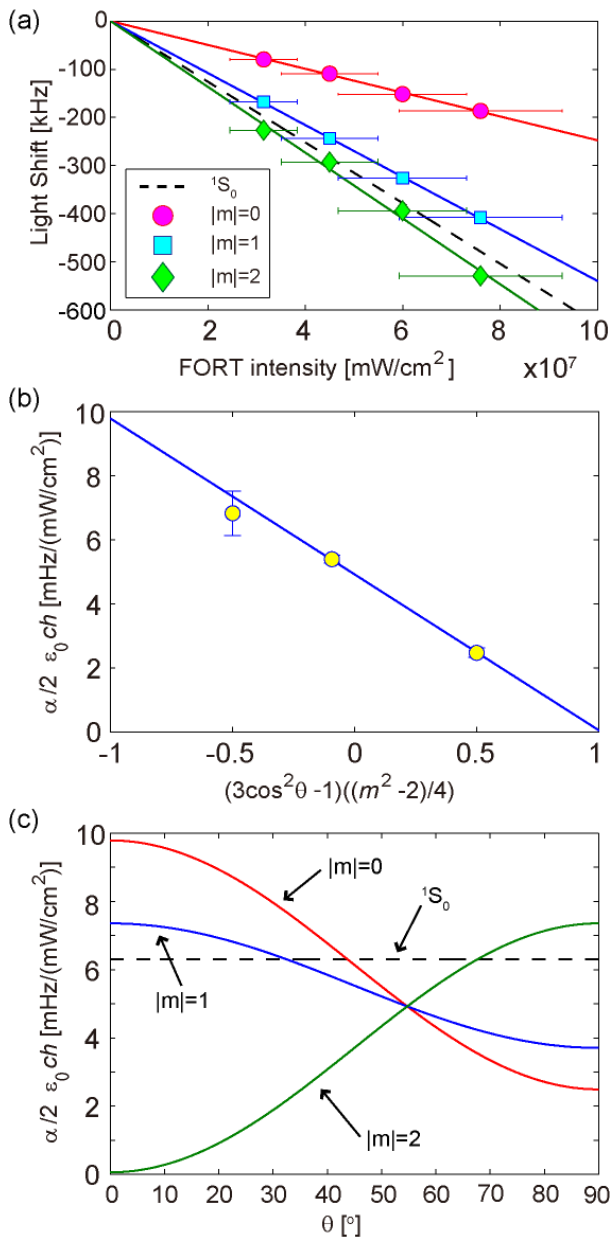} %2column
\caption{(Color online)
(a) AC Stark shift of the $^3$P$_2$ state of $^{174}$Yb in a laser field with a wavelength of 1070 nm.
The horizontal axis is the intensity of the horizontal FORT beam.
The red circles, blue squares, and green diamonds correspond to the data for 
$(m,\theta) = (0,89^{\circ})$, $(-1,47^{\circ})$, and $(-2,89^{\circ})$.
Solid lines are the linear fits to the data.
The shift of the $^1$S$_0$ state is also plotted as a black dashed line for reference.
(b) The polarizability of the $^3$P$_2$ state is plotted as a function of $(3\cos^2\theta-1)((m^{2}-2)/4)$.
Three data points correspond to 
$(m,\theta) = (0,89^{\circ})$, $(-1,47^{\circ})$, and $(-2,89^{\circ})$ from left.
The solid line is the linear fit based on Eq. (\ref{eq:polarizability}).
(c) $\theta$-dependence of the polarizabilities of the $^3$P$_2$ states.
The polarizabilities are evaluated by using the experimentally determined scalar and tensor polarizabilities.
The red, blue, and green lines correspond to $|m|=$ 0, 1, and 2.
These values coincide at the condition where $\cos{{^2}{\theta}=1/3}$ 
as can be seen from Eq. (\ref{eq:polarizability}).
The polarizability of the $^1$S$_0$ state is also plotted for reference.
The intersection point with the black dashed line indicates the "magic wavelength" condition for each sublevel.}
\label{fig:polarizability}
\end{figure}

In the experiment, we measure the transition frequencies of the $^1$S$_0$-$^3$P$_2$ ($m = 0, -1, -2$) in the crossed FORT 
with changing the power of the horizontal FORT beam with a wavelength of 1070 nm.
The polarization of the horizontal FORT is on the x-y plane.
The excitation beam propagates along the y-axis and its polarization is parallel to the z-axis.
In order to satisfy the selection rule for the $^1$S$_0$-$^3$P$_2$ transition, 
the magnetic field of about 1 G is applied 
along the direction parallel to the horizontal FORT beam, 
the x direction, and the z direction for the excitation to the states of $m=0$, $m=-1$, and $m = -2$, respectively.
The angle $\theta$ between the polarization of the laser field and the magnetic field is 
$89^{\circ}\pm2^{\circ}$, $47^{\circ}\pm1^{\circ}$, and $89^{\circ}\pm2^{\circ}$ for $m=0$, $m=-1$, and $m=-2$, respectively.
In our condition the direction of the magnetic field sets a quantization axis.
After evaporative cooling we irradiate the excitation laser for 50 ms in the case of $m = -2$ and for 100 ms in the case of $m = 0$ and $-1$.
We measure the number of $^{174}$Yb atoms remaining in a trap with changing the frequency of the excitation laser.
For each sublevel $m$, we evaluate the amount of AC Stark shift of the $^3$P$_2$ state 
by subtracting the shift of the $^1$S$_0$ state from the resonant frequencies in Fig. \ref{fig:polarizability} (a).
The calculated light shift of the $^1$S$_0$ state is also plotted for reference.
The polarizabilities of the $^3$P$_2$ state measured in this setup are listed in Table \ref{tab:polarizability}.
We show $\alpha_{e}/2\epsilon_{0}ch$ instead of $\alpha_e$ itself for convenience.

\begin{table}
\caption{The polarizabilities of the $^3$P$_2$ state $\alpha/2\epsilon_{0}ch$ in the laser field with a wavelength of 1070 nm.
$\theta$ is an angle between the polarization of the laser field and the quantization axis.
The corresponding value of the $^1$S$_0$ state is also added to this table for reference.}
\label{tab:polarizability}
\begin{center}
% \begin{tabular}{ccc} \toprule 
%  $m$ & $\theta$ & $\alpha/2\epsilon_{0}ch\mathrm{[mHz/(mW/cm^2)]}$ \\ \midrule
%  $0$ & $89^{\circ}\pm2^{\circ}$ & $2.5 \pm 0.1$ \\
%  $-1$ & $47^{\circ}\pm1^{\circ}$ & $5.4 \pm 0.1$ \\
%  $-2$ & $89^{\circ}\pm2^{\circ}$ & $6.8 \pm 0.7$ \\
%  ($^1$S$_0$) &  & 6.3 \\ \bottomrule
% \end{tabular}
\begin{tabular}{ccc} \hline 
 $m$ & $\theta$ & $\alpha/2\epsilon_{0}ch\mathrm{[mHz/(mW/cm^2)]}$ \\ \hline
 $0$ & $89^{\circ}\pm2^{\circ}$ & $2.5 \pm 0.1$ \\
 $-1$ & $47^{\circ}\pm1^{\circ}$ & $5.4 \pm 0.1$ \\
 $-2$ & $89^{\circ}\pm2^{\circ}$ & $6.8 \pm 0.7$ \\
 ($^1$S$_0$) &  & 6.3 \\ \hline
\end{tabular}
\end{center}
\end{table}

\begin{table}
\caption{The angles $\theta$ that meets the equal polarizability condition for the transition $^1$S$_0$ - $^3$P$_2$ in the laser field with $\lambda = 1070$ nm.}
\label{tab:magic}
\begin{center}
% \begin{tabular}{cc} \toprule 
%  $m$ & $\theta$ \\ \midrule
%  $0$ & $43.7^{\circ} \pm 1.5^{\circ}$ \\
%  $\pm1$ & $32.5^{\circ} \pm 3.3^{\circ}$ \\
%  $\pm2$ & $67.7^{\circ} \pm 2.1^{\circ}$ \\ \bottomrule
% \end{tabular}
\begin{tabular}{cc} \hline
 $m$ & $\theta$ \\ \hline
 $0$ & $43.7^{\circ} \pm 1.5^{\circ}$ \\
 $\pm1$ & $32.5^{\circ} \pm 3.3^{\circ}$ \\
 $\pm2$ & $67.7^{\circ} \pm 2.1^{\circ}$ \\ \hline
\end{tabular}
\end{center}
\end{table}

In the case of linearly polarized light, the polarizability of the $^3$P$_2$ state is described as \cite{polarizability}
\begin{equation}
\alpha_{e} = \alpha^{s}_{nJ}+(3\cos^2\theta-1)\alpha^T_{nJ}\frac{3m^{2}-J(J+1)}{2J(2J-1)} .
\label{eq:polarizability}
\end{equation}
Here $\alpha^s_{nJ}$ is a scalar polarizability, $\alpha^T_{nJ}$ is a tensor polarizability, 
$J = 2$ is the total angular momentum of the atom for the $^3$P$_2$ state of $^{174}$Yb, 
and $n$ is the set of the remaining quantum numbers.
Therefore, $\alpha_{e}$ is a linear function of $(3\cos^2\theta-1)((m^{2}-2)/4)$ 
as represented in Fig. \ref{fig:polarizability} (b). 
From the linear fit based on Eq. (\ref{eq:polarizability}), the scalar and the tensor polarizabilities are 
determined as $\alpha^s_{nJ}/2\epsilon_{0}ch=4.9\pm0.1$ and $\alpha^T_{nJ}/2\epsilon_{0}ch=-4.9\pm0.4$ mHz/(mW/cm$^2$), respectively.
In Fig. \ref{fig:polarizability} (c) we plot the polarizability of each sublevel $m$ as a function of the angle $\theta$ between the laser polarization and the magnetic field.
It is noted that the polarizability is positive.
From these values, all the magnetic sublevels of the $^3$P$_2$ state of $^{174}$Yb are trappable by a laser field with wavelength of 1070 nm.
In addition, if we adjust the angle $\theta$ for each magnetic sublevel, we can satisfy $\alpha_{g} = \alpha_{e}$, like a magic wavelength condition.
Each value of $\theta$ that meets the condition of the "magic wavelength" is listed in Table \ref{tab:magic}.

The polarizability of a state $|\mathrm{a}\rangle$ with an energy $E_{\mathrm{a}}=\hbar\omega_{\mathrm{a}}$ is given by the following formula.
\begin{multline}
\frac{1}{2\epsilon_{0}c}
\alpha_{\mathrm{a}} = 
\sum_{\mathrm{b}\neq\mathrm{a}} 
\frac{e^{2}}{\epsilon_{0}\hbar c}
\begin{pmatrix}
J_{\mathrm{b}} & 1 & J_{\mathrm{a}} \\
-M_{\mathrm{b}} & 0 & M_{\mathrm{a}}
\end{pmatrix}
^{2} \\
\times |\langle n_{\mathrm{b}}L_{\mathrm{b}}J_{\mathrm{b}}\|r\|n_{\mathrm{a}}L_{\mathrm{a}}J_{\mathrm{a}} \rangle|^2
\frac{\omega_{\mathrm{b}}-\omega_{\mathrm{a}}}{(\omega_{\mathrm{b}}-\omega_{\mathrm{a}})^2-{\omega}^2} ,
\label{eq:perturbation}
\end{multline}
where $\hbar \omega_{\mathrm{b}} = E_{\mathrm{b}}$ is an energy of another state $|\mathrm{b}\rangle$ 
and the sum runs over all the states which can be coupled with the state $|\mathrm{a}\rangle$ through electric dipole transitions.
Here 
$\bigl( \begin{smallmatrix}
J_{\mathrm{b}} & 1 & J_{\mathrm{a}} \\
-M_{\mathrm{b}} & 0 & M_{\mathrm{a}}
\end{smallmatrix} \bigr)$, 
$|\langle n_{\mathrm{b}}L_{\mathrm{b}}J_{\mathrm{b}}\|r\|n_{\mathrm{a}}L_{\mathrm{a}}J_{\mathrm{a}} \rangle|$, and $\omega$ denote 
the Wigner 3-$j$ symbol, the reduced matrix element, and the laser frequency, respectively.
$L$, $J$, $M$, and $n$ are the orbital angular momentum of electrons, 
the total angular momentum of electrons, its projection onto the quantization axis, 
and the set of the remaining quantum numbers.
From the theoretically known values of the reduced matrix elements for (5d6s)$^3$D$_1$, (5d6s)$^3$D$_2$, (5d6s)$^3$D$_3$, (5d6s)$^1$D$_2$, and (6s7s)$^3$S$_1$ \cite{RME}, 
we obtain $\alpha(m=0)/2\epsilon_{0}ch = 4.6$, $\alpha(|m|=1)/2\epsilon_{0}ch = 2.2$, and $\alpha(|m|=2)/2\epsilon_{0}ch =-5.0$ mHz/(mW/cm$^2$), respectively.
On the other hand, 
the experimental values of polarizabilities $\alpha/2\epsilon_{0}ch$ at $\theta = 0^{\circ}$ are $9.8\pm0.5$, $7.4\pm0.3$, and $0.06\pm0.47$ mHz/(mW/cm$^2$) 
for $|m| = 0, 1$, and $2$, respectively.
The theoretical values are, therefore, smaller than the experimental ones for all $m$.
These discrepancies would originate from the lack of information of higher energy states, 
and the fact that all the higher energy states make a contribution of positive values to the polarizabilities.
As an attempt to resolve the discrepancy we consider only the contribution from the next higher energy states.
Then it is found that the reduced matrix elements with the values of 
$|\langle \mathrm{(5d6s^2)^3D_1}\|r\| ^3\mathrm{P}_2 \rangle|$ = $|\langle \mathrm{(6s6d)^3D_3}\|r\| ^3\mathrm{P}_2 \rangle|$ = 2.8 (a.u.) 
and $|\langle \mathrm{(6s6d)^3D_1}\|r\| ^3\mathrm{P}_2 \rangle|$ = $|\langle \mathrm{(6s6d)^3D_2}\|r\| ^3\mathrm{P}_2 \rangle|$ = 4.6 (a.u.), 
can bring the theoretical values close to the experimentally obtained values as 
$\alpha(m=0)/2\epsilon_{0}ch=9.8$, $\alpha(|m|=1)/2\epsilon_{0}ch =7.4$, and $\alpha(|m|=2)/2\epsilon_{0}ch=-0.05$ mHz/(mW/cm$^2$).

Here we discuss the wavelength dependence on the polarizability.
If the wavelength differs by less than 13 nm, 
the variation of each term in Eq. (\ref{eq:perturbation}) is within 10\%.
Due to this small variation we consider that the polarizabilities of the vertical FORT (1083 nm) and lattice beams (1064 nm) 
are almost the same as those of the horizontal FORT (1070 nm).

\subsection{High resolution spectroscopy of $^{174}$Yb atoms in an optical lattice}

\begin{figure}
\centering
\includegraphics[width=7cm,keepaspectratio]{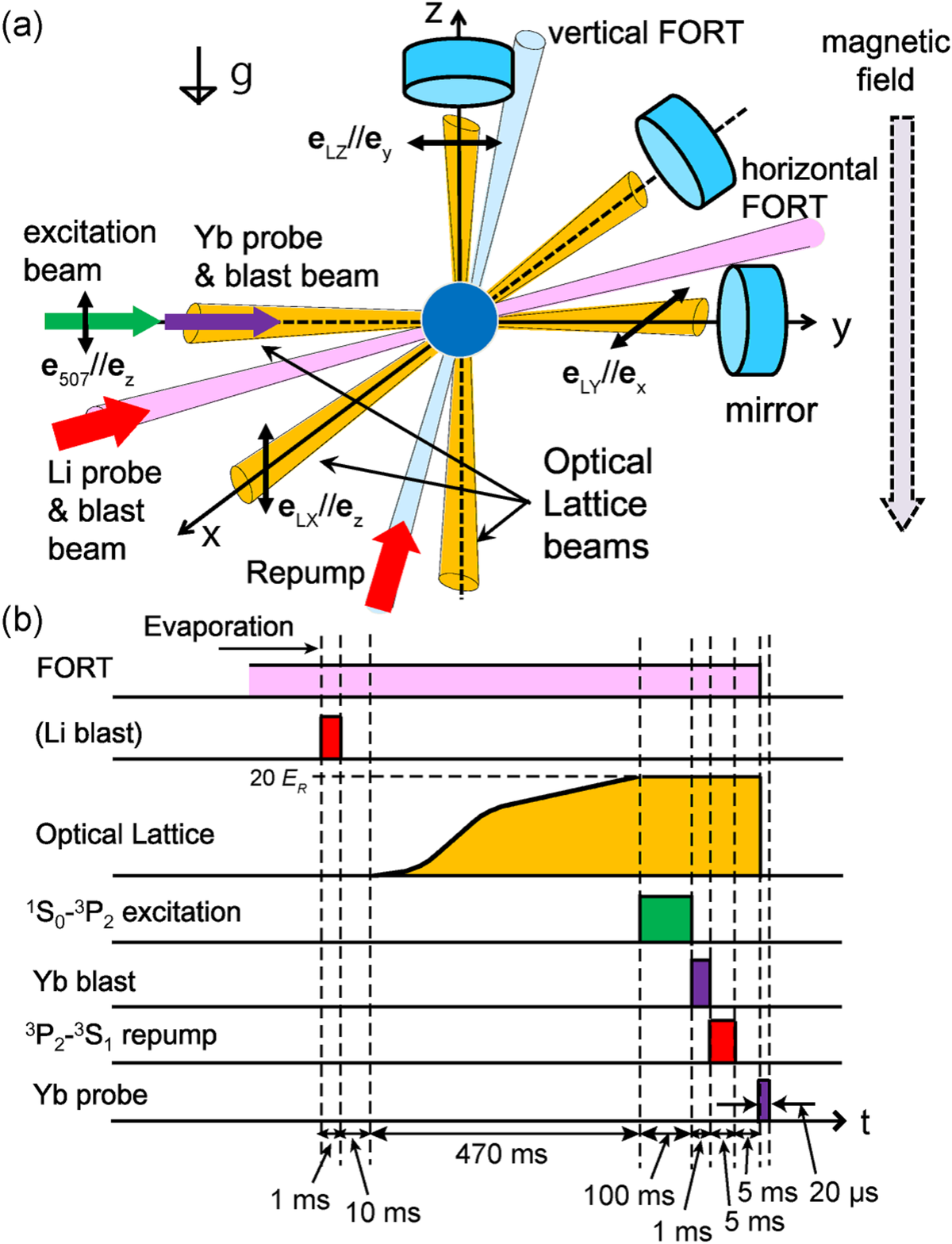}
\caption{(Color online)
Experimental setup (a) and sequence (b) to perform a spectroscopy of $^{174}$Yb in an optical lattice.
For the pure $^{174}$Yb we remove $^6$Li by a blast laser.
For the mixture we wait for the same duration without irradiating the Li blast laser.
$\mathbf{e}_{507}$, $\mathbf{e}_{LX}$, $\mathbf{e}_{LY}$, and $\mathbf{e}_{LZ}$ are the unit polarization vectors of the excitation laser, x, y, and z-lattice.
They are parallel to the z, z, x, and y-axis respectively.
The external magnetic field is applied along the z-axis.}
\label{fig:seuptimechartspectroscopy}
\end{figure}

We perform a laser spectroscopy of $^{174}$Yb atoms in an optical lattice using the ultra-narrow optical transition $^1$S$_0$-$^3$P$_2$ 
with and without the Fermi sea of $^6$Li.
Figure \ref{fig:seuptimechartspectroscopy} (a) shows the experimental setup.
In this experiment we excite the atoms to the magnetic sublevel $ m = -2$ 
because the $m = -2$ state does not suffer from inelastic decay due to Zeeman sublevel changes in the $^1$S$_0$-$^3$P$_2$ collisions \cite{3P2collision}.
The external magnetic field is applied along the z-axis, 
and the polarizations of FORT lasers and the y, z-lattice beams are perpendicular to the magnetic field.
Even if the lattice potential for the $^1$S$_0$ state is isotropic, 
the potential for the $^3$P$_2$ is anisotropic due to $\theta$-dependence.
When the depth for the $^1$S$_0$ state $^{174}$Yb is 20 $E_{R}^{\mathrm{Yb}}$, 
the potential depths for the $^3$P$_2$ ($m=-2$) state created by the x, y, and z-lattices are 
$(0.2\pm1.5)$, $(23.3\pm0.8)$, and $(23.3\pm0.8)$ $E_{R}^{\mathrm{Yb}}$, respectively.
This indicates that the potential for the $^3$P$_2$ state is like the arrays of one-dimensional tubes in this condition.

The detail of the experimental procedure is shown in Fig. \ref{fig:seuptimechartspectroscopy} (b).
After ramping up the lattice potential depth to 20 $E_{R}^{\mathrm{Yb}}$ in the same manner as the visibility measurement, 
a portion of $^{174}$Yb atoms in the $^1$S$_0$ state are excited to the $^3$P$_2$ ($ m = -2$) state for 100 ms.
The excitation laser propagates along the y-axis as shown in Fig. \ref{fig:seuptimechartspectroscopy} (a).
The step of the frequency scan is 4 kHz.
The remaining $^{174}$Yb atoms in the $^1$S$_0$ state are removed from the trap by a laser pulse resonant to the $^1$S$_0$-$^1$P$_1$ transition (Yb blast laser, $\lambda$ = 399 nm) for 1 ms.
Then, the atoms in the $^3$P$_2$ state are repumped through the $^3$S$_1$ state 
by irradiating a laser pulse ($\lambda = 770$ nm) for 5 ms.
See Fig. \ref{fig:energydiagram} for relevant energy levels.
In this experiment, another repumping laser resonant to the $^3$P$_0$-$^3$S$_1$ transition is not used.
After 5 ms holding time, all the trap potentials and the external magnetic field are turned off and the absorption image is taken without any expansion time.

\begin{figure}
\centering
\includegraphics[width=5.5cm,keepaspectratio]{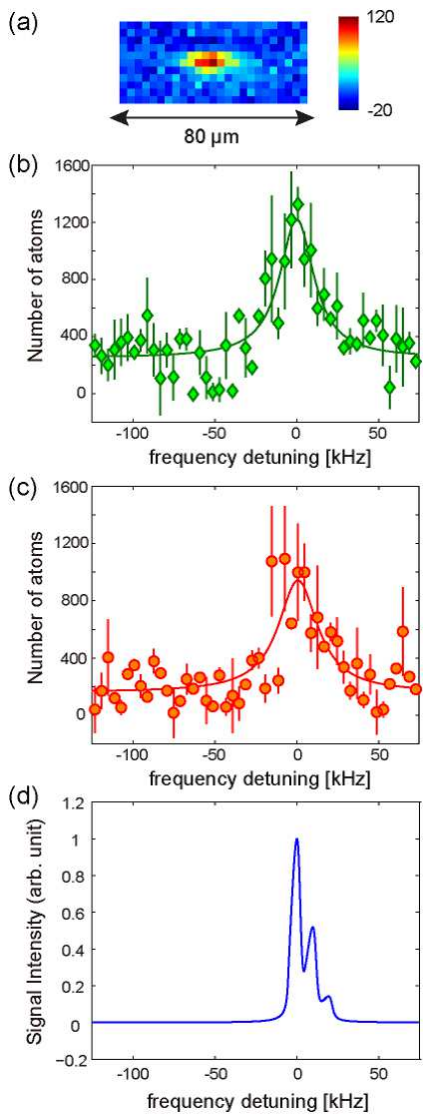} %2column
\caption{(Color online)
Results of the laser spectroscopy using the ultra-narrow transition of the $^1$S$_0$-$^3$P$_2$ ($m = -2$).
(a) A typical absorption image of the $^{174}$Yb atoms repumped back to the $^1$S$_0$ state.
The color bar indicates the number of the $^{174}$Yb atoms in each pixel.
(b) Excitation spectrum for the pure $^{174}$Yb, 
where the number of repumped atoms are plotted as a function of the excitation laser frequency.
The error bars indicate the standard errors.
The curve line is the unweighted Lorentz fit to the data.
The origin of the frequency detuning is set at the center frequency of the Lorentz fit.
(c) Excitation spectrum for the mixture.
The curve is the unweighted Lorentz fits to the data.
The origin of the frequency detuning is the same as that for the pure $^{174}$Yb case.
(d) Calculated spectrum at the atomic limit for the pure $^{174}$Yb.}
\label{fig:spectroscopy}
\end{figure}

Figure \ref{fig:spectroscopy} (a) is a typical absorption image of the $^{174}$Yb atoms returned back to the $^1$S$_0$ state.
We evaluate the number of atoms after subtracting the background level by using the surrounding area of the image.
The excitation laser is generated 
by the frequency doubling of a laser diode with a wavelength of 1014 nm.
The laser diode is frequency locked to a high-finesse ultralow-expansion cavity with the long-term drift of 1.4 kHz/hour.
In order to compare the difference between the cases with and without $^6$Li atoms at the same frequency, 
Li blast laser is alternatively turned on and off sequentially.
Figure \ref{fig:spectroscopy} (b) and \ref{fig:spectroscopy} (c) shows 
the obtained excitation spectra for the pure $^{174}$Yb and the mixture, respectively, 
where the number of atoms repumped back to the $^1$S$_0$ state is plotted.
Each data point represents the average over three data and the error bars indicate standard errors.

Here we discuss the excitation spectra for the pure $^{174}$Yb case.
The spectrum shows a single broad peak.
The broadening can be explained as follows.
Firstly, it should include resonances from the multiply occupied sites.
The frequencies corresponding to the resonances from the multiply occupied sites $n$, are shifted by $(n-1) \times (U_{ge}-U_{gg})/h$ 
relative to the resonance of the singly occupied sites.
Here $U_{ge}$ is the on-site interaction between the $^1$S$_0$ and the $^3$P$_2$ states 
and $U_{gg}$ is that between the $^1$S$_0$ states.
When the potential for the $^1$S$_0$ state is 20 $E_{R}^{\mathrm{Yb}}$, 
the interaction strengths are $U_{ge}/h=-0.15$ kHz and $U_{gg}/h=0.44$ kHz.
Thus the frequency differences between the neighboring peaks are 0.60 kHz, 
and the peaks related to multiply occupied sites are located at the negative side of the peak from singly occupied sites.
According to the estimation in the atomic limit $U/zJ\to\infty$, 
the relevant peaks exist over a range of 7.3 kHz corresponding to the occupation numbers of up to $n=11$.
Secondly, the excitation spectra include resonances to the higher vibrational states in the optical lattice on the positive frequency side.
Since the Lamb-Dicke parameter \cite{winelanditano} takes a relatively large value of 0.7, 
the ratio of the transition strength of the carrier, the first, and the second sidebands is $1:0.5:0.1$.
The frequency spacing is estimated to be 9.8 kHz.
Therefore the peaks related to the excitations to higher vibrational states exist over a range of 20 kHz.
Thirdly, each peak should be broadened up to 3 kHz by the inhomogeneity of the potential.
For these reasons the width of the excitation spectrum is about 28 kHz, 
where this width is defined by the difference of the frequency detunings at which the signal intensity is equal to 10 \% of the peak value.
In Fig. \ref{fig:spectroscopy} (d) we show the calculated spectrum at the atomic limit.
Apart from the clear structures, which are not observed in the experiments, this result is almost consistent with the obtained spectrum.

The observed spectra do not show clear difference between the cases with and without $^6$Li.
Possible reasons for this small difference are the small spatial overlap and the small interspecies interaction strength, 
which are similarly considered as the origin of the small difference of visibility.
This experiment is, however, an important first step towards the exploration of time-dependent impurity problems such as Anderson's Orthogonality Catastrophe, 
which could be studied 
if we find a Feshbach resonance between the $^2$S$_{1/2}$ state of $^6$Li atoms and the $^3$P$_2$ state of $^{174}$Yb atoms, 
possibly introduced by anisotropy induced Feshbach resonance, as in the $^1$S$_0$ (Yb) and $^3$P$_2$ (Yb) case \cite{3P2Feshbach}.

\section{Conclusion}
In conclusion, we develop an optical lattice system for an ultracold atom mixture of $^{174}$Yb and $^6$Li.
We confirm the successful formation and the loading of the optical lattice 
by observing the matter-wave interference patterns of $^{174}$Yb atoms at a relatively shallow optical lattice.
We also perform a high-resolution laser spectroscopy of the $^{174}$Yb atom in the presence of the Fermi sea of $^6$Li.
Although the clear differences in the interference patterns and the excitation spectra between the mixture and the pure $^{174}$Yb are not observed, 
these experiments are important first steps towards the research of impurity problems.
In addition we measure the polarizabilities of the $^3$P$_2$ state $^{174}$Yb in an optical trap with a wavelength of 1070 nm and 
the scalar and tensor polarizabilities are determined.
It is found that the polarizability of the $^3$P$_2$ state $^{174}$Yb with a wavelength around 1070 nm 
can take a various value by properly tuning the angle between the polarization and the external magnetic field 
and by selecting the magnetic substate.

\begin{acknowledgments}
%\acknowledgment
This work was supported by the Grant-in-Aid for Scientific Research of JSPS (No. 18204035, 25220711 , 21102005C01 ( Quantum Cybernetics), 21104513A03 (DYCE), 22684022), 
GCOE Program "The Next Generation of Physics, Spun from Universality and Emergence" from MEXT of Japan, 
World-Leading Innovative R$\&$D on Science and Technology (FIRST), and Matsuo Foundation.
H. H. , H. K. and S. N. acknowledge support from JSPS.
\end{acknowledgments}

\end{document}